\documentclass[12pt]{iopart}

\usepackage{graphicx}
\usepackage{epsfig}
\usepackage{color}
\usepackage{iopams}
\usepackage{hyperref}
\usepackage[latin1]{inputenc}

\begin{document}
\title{Electron Spin Resonance of the ferromagnetic Kondo lattice CeRuPO}

\author{T F\"orster, J Sichelschmidt, C Krellner, C Geibel, and F Steglich}

\address{Max Planck Institute for Chemical Physics of Solids, D-01187 Dresden, Germany}

\begin{abstract}
The spin dynamics of the ferromagnetic Kondo lattice CeRuPO is investigated by Electron Spin Resonance (ESR) at microwave frequencies of 1, 9.4, and 34~GHz. The measured resonance can be ascribed to a rarely observed bulk Ce$^{3+}$ resonance in a metallic Ce compound and can be followed below the ferromagnetic transition temperature $T_C=14$~K. At $T>T_C$ the interplay between the RKKY-exchange interaction and the crystal electric field anisotropy determines the ESR parameters. Near $T_C$ the spin relaxation rate is influenced by the critical fluctuations of the order parameter.
\end{abstract}


\section{Introduction}

Until the discovery of an Electron Spin Resonance (ESR) signal below the Kondo temperature $T_K$ in YbRh$_2$Si$_2$ \cite{Sich03} it was believed and experimentally well manifested that there should be no ESR signal in heavy fermion systems. This was justified by the strong electronic correlations which originate from the hybridization of the 4$f$/5$f$-electrons and the conduction electrons \cite{Woelf09}. Since 2003 only two other Kondo lattices have been found which show an ESR Signal with local properties of their Kondo-ions namely YbIr$_2$Si$_2$ (I-type) \cite{Sich07} and CeRuPO \cite{Krel08}.

In CeRuPO a pronounced decrease of the electrical resistivity at temperatures below 50~K indicates the onset of coherent Kondo scattering. Ferromagnetic (FM) order appears at $T_C$ = 14~K and therefore the system is a rare case of a FM Kondo lattice. In contrast to the Yb-compounds mentioned above, the Ruderman-Kittel-Kasuya-Yosida (RKKY) interaction is stronger than the Kondo interaction characterized by a Kondo temperature $T_K\approx10$~K \cite{Krel07a}. It was a remarkable observation that in contrast to its antiferromagnetic homologue CeOsPO the ferromagnetic CeRuPO shows a well-defined signal in the regime of coherent Kondo scattering \cite{Krel08}. This fact suggests that FM correlations are important for the observability of a Kondo-ion ESR, which was later on supported by several theoretical investigations \cite{Woelf09,Schl09,Koch09,Zvya09a}. 

This paper reports detailed ESR investigations of CeRuPO in order to document the properties of the rarely observed Ce$^{3+}$ spin resonance (we are aware of only two others, namely CeP \cite{Huan76} and CeB$_6$ \cite{Demi09a}), and to show the effect of ferromagnetic correlations and magnetic order on the Kondo ion spin resonance. 

The system reveals a unique magnetic anisotropy: Although CeRuPO is a collinear ferromagnet with the magnetic moments aligned along the $c$-axis, the saturation magnetization along this axis is smaller than in the basal plane. This behaviour originates from different anisotropies of the crystal electric field (CEF) and the RKKY exchange interaction with respect to the components of the magnetic moment. The first one results in a ground state CEF doublet with a larger saturation moment in the basal plane. The latter favours an alignment of the moments along the $c$-axis and is responsible for the FM transition \cite{Krel08a}. 

\section{Experimental Details}

The ESR experiments were carried out using a standard continuous wave spectrometer together with a He-flow cryostat that allows us to vary the temperature from 4 to 300~K. In order to investigate the magnetic field dependence we used three frequencies $\nu=1, 9.4$, and $34$~GHz (L-, X-, Q-band); for a $g$-factor of 2 this corresponds to resonance fields of 36, 340, and 1200~mT. For the lowest frequency we used a split-ring resonator, which has a lower Q factor than the resonant cavities utilized at higher frequencies.

ESR probes the absorbed power $P$ of a transversal magnetic microwave field as a function of a static and external magnetic field $\mu_0H$. To improve the signal-to-noise ratio, we used a lock-in technique by modulating the static field, which yields the derivative of the resonance signal $dP/dH$. The measured ESR spectra were fitted with a metallic Lorentzian function including the influence of the counter-rotating component of the linearly polarized microwave field \cite{Wykh07a}. From the fit we obtained the resonance field $H_{res}$ (which determines the ESR $g$-factor $g=h\nu/\mu_BH_{res}$), the ESR intensity $I_{ESR}$, and the linewidth $\Delta H$ (half-width at half maximum). In metals $\Delta H$ is a direct measure of the spin lattice relaxation time $1/T_1$ and its temperature and frequency dependence reveals the nature of the participating relaxation mechanisms.

For our measurements we used three CeRuPO single crystals (one for each frequency) which were grown from the same batch by a Sn-flux method \cite{Krel08a}. CeRuPO crystallizes in the tetragonal ZrCuSiAs-type structure ($P4/nmm$) containing alternating layers of OCe$_4$ and RuP$_4$ tetrahedra. The platelet-like single crystals had a surface area of about 3~mm$^2$ and a thickness of up to 0.08~mm with the crystallographic $c$-axis perpendicular to the platelet plane. Electron-microprobe, X-ray diffraction, and a large residual resistivity ratio ($\rho_{300K}/\rho_0=30$) indicated a high sample quality of single-phase CeRuPO \cite{Krel08a}.

\begin{figure}[t]
 \centering
\includegraphics[width=0.5\textwidth]{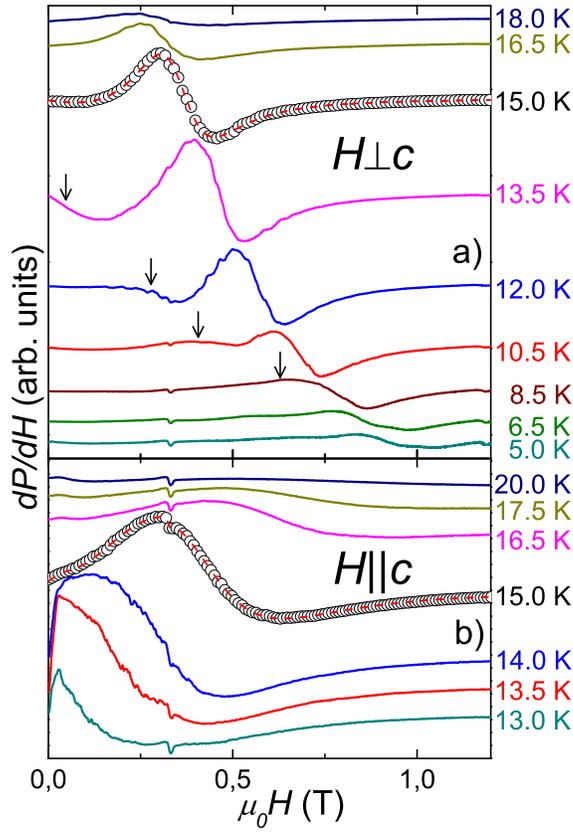}
 \caption{Typical ESR spectra at 9.4~GHz of single crystalline CeRuPO at different temperatures and field orientations. Dashed lines represent fits with a single metallic Lorentzian. Arrows in a) indicate structures which occur below $T_C$. The structure at 0.33~T originates from the background.}
\label{signals}
\end{figure}
\section{Experimental Results and Discussion}

Figure \ref{signals} shows the temperature evolution of the ESR spectra for two orientations of the crystallographic $c$-axis: perpendicular ($H\bot c$) and parallel ($H\|c$) to the quasistatic magnetic field $\mu_0H$ at 9.4~GHz. In the paramagnetic region the ESR signal could be observed in a small temperature range up to $\approx 23$~K, where it is in good agreement with a single metallic Lorentzian shape for both orientations (dashed lines in \autoref{signals}). In the ferromagnetically ordered region, the signal displays strongly anisotropic properties: for $H\bot c$ additional structures below the main line appear (see arrows in \autoref{signals}a) which probably can be related to ferromagnetic resonance modes. The resonance field of the main line increases with lowering the temperature and the signal was visible down to at least 4~K. For $H\|c$ the ESR line rapidly broadens with decreasing temperature and strongly shifts towards a zero resonance field. Therefore, at $H\|c$, the signal could be detected down to 13~K for X-band and 9~K for Q-band. For the less sensitive L-band setup the signal could only be observed for $H\bot c$ between 12 and 15.5~K. Below we only discuss the ESR parameters which were obtained by fitting the spectra with Lorentzian lines that describe linewidth and resonance field within an error of 15\%. 

The bulk origin of the ESR signal is corroborated by its large signal-to-noise ratio and especially by the angle and the temperature dependences (see following sections). Also the ESR intensity follows the magnetic susceptibility $\chi$, and we find a linear relation $I_{ESR}\propto(\chi-\chi^0)$ with $\chi_{\bot}^0$=0.09$\cdot$10$^{-6}$~m$^3$/mol and $\chi_{\|}^0$= 1.0$\cdot$10$^{-6}$~m$^3$/mol (not shown) similar to what was found in YbRh$_2$Si$_2$ \cite{Sich03}. A quantitative comparison of the ESR intensity with a CuSO$_4$ standard confirms that all Ce$^{3+}$ ions within the penetration depth of the microwave contribute to the signal.

In our previous publication on polycrystalline CeRuPO we reported a well-defined ESR signal which splits into a low-field (LF) and a high-field (HF) component near $T_C$ (note that in the polycrystalline samples $T_C=15$~K). Both signals showed a pronounced temperature behaviour of their resonance fields which shift in opposite directions. We suspected that the components belong to different orientations of the powder grains crystal axes to the resonance field \cite{Krel08}. The single crystal data presented here confirms this assumption and we could identify the LF and HF component with powder grains oriented around a direction parallel to $H\|c$ and $H\bot c$.


The strong shift of the resonance field below the ferromagnetic ordering temperature indicates the relevance of demagnetization and magnetic anisotropy fields. These effects strongly depend on the orientation of the external field with respect to the magnetic- and shape-anisotropy of the sample. The magnetic easy axis coincides with the crystallographic $c$-axis. Then, in the ferromagnetic phase and $H\| c$ the resonance condition reads:
\begin{equation}
h\nu=g\mu_B \left(H_{res}-NM(H,T)+H_A(T)\right),
\label{FMRhpara}
\end{equation}
where $M(H,T)$ is the magnetization of the sample and $H_A$ is the anisotropy field. In order to calculate demagnetizing effects the magnetization $M(H,T)$ was measured for $H\| c$ in a separate experiment using a commercial Quantum Design SQUID magnetometer. The samples are thin platelets, and therefore, we assume a demagnetizing factor of $N=1$ for the magnetization normal to the platelet plane ($H\| c$) and of $N=0$ for the perpendicular direction ($H\bot c$). It turns out that in the temperature region we discuss here (12-14~K) $NM(H,T)$ is smaller than 1\% of the measured resonance field and, thus, $NM(H,T)$ can safely be neglected. Using \autoref{FMRhpara} with $g_{\|}=$~1.18 and the 34~GHz data we find for the anisotropy field $\mu_0H_A\approx$~1.4~T at 10~K and $\mu_0H_A\approx$~0.65~T at 12~K. Therefore, the temperature dependence of the anisotropy field easily exceeds the estimated 20\% temperature variation of the $g$-factor anisotropy: $g_\|/g_\perp=0.46$ at 20 K; $g_\|/g_\perp=M_\|/M_\perp=0.36$ at 2 K, Ref.\cite{Krel08a}. A more detailed analysis of the behaviour below $T_C$ as it was done for FM thin films in Refs. \cite{Anis99} and \cite{Farl00} does not lead to reasonable results for CeRuPO. The huge linewidth and the metallic lineshape prevents a sufficent accuracy for the determination of the resonance field. Therefore, most of our conclusions are only valid above $T_C$. Although the effect of ferromagnetism regarding its anisotropy fields vanishes at $T_C$ critical fluctuations may still influence the temperature dependence of the linewidth $\Delta H$ \cite{Kotz75}.

\begin{figure}[t]
 \centering
\includegraphics[width=0.75\textwidth]{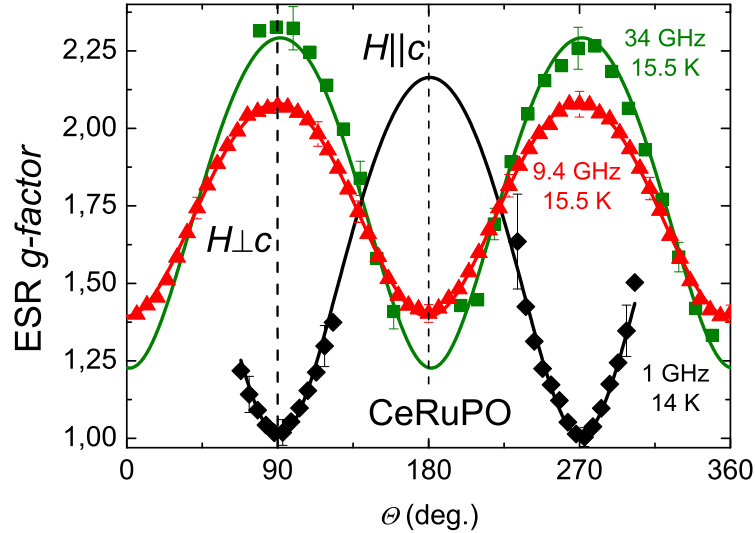}
 \caption{Anisotropy of the effective g-factor at different frequencies (1~GHz 14~K \fulldiamond, 9.4 GHz 15.5~K \textcolor{red}{\fulltriangle}, 34~GHz 15.5~K \textcolor[rgb]{0,0.5,0}{\fullsquare}).  $\Theta$ denotes the angle between the field $H$ and the $c$-axis. Solid lines represent fits a with a uniaxial symmetry (see equation (\ref{gunixial})). The 1~GHz data are plotted at a temperature where $g(\Theta)$ is phase-shifted by $90^\circ$ compared to the behaviour above 15~K.}
\label{gtheta}
\end{figure}

\subsection{Anisotropy and temperature dependence of the g-factor}
Figure \ref{gtheta} presents the angle dependence of the effective ESR g-factor at the different frequencies. The crystals were rotated by an angle $\Theta$ around an axis in the basal plane. The data can nicely be fitted with an uniaxial symmetry behaviour (solid lines in \autoref{gtheta}): 
\begin{equation}
g(\Theta)= \sqrt{g^2_{\|}\cos^2\Theta+g^2_{\bot}\sin^2\Theta}.
\label{gunixial}
\end{equation}
This is expected for a tetragonal crystal structure and could furthermore be confirmed by rotating the crystal around the c-axis, keeping $\Theta=90^\circ$ constant: within experimental accuracy all ESR parameters were found to be isotropic in this configuration.

The $g$-factor anisotropy reveals a remarkable temperature dependence, namely a crossing of $g_{\|}(T)$ and $g_{\perp}(T)$ resulting in a $90^\circ$-phase-shift of $g(\Theta)$ at a temperature which depends on the magnetic field. In \autoref{gtheta} the phase shift can be seen for the 1~GHz data compared to the data at 9.4 and 34~GHz which are shown for temperatures where the phase shift has not yet occurred.

\begin{figure}[t]
 \centering
\includegraphics[width=0.6\textwidth]{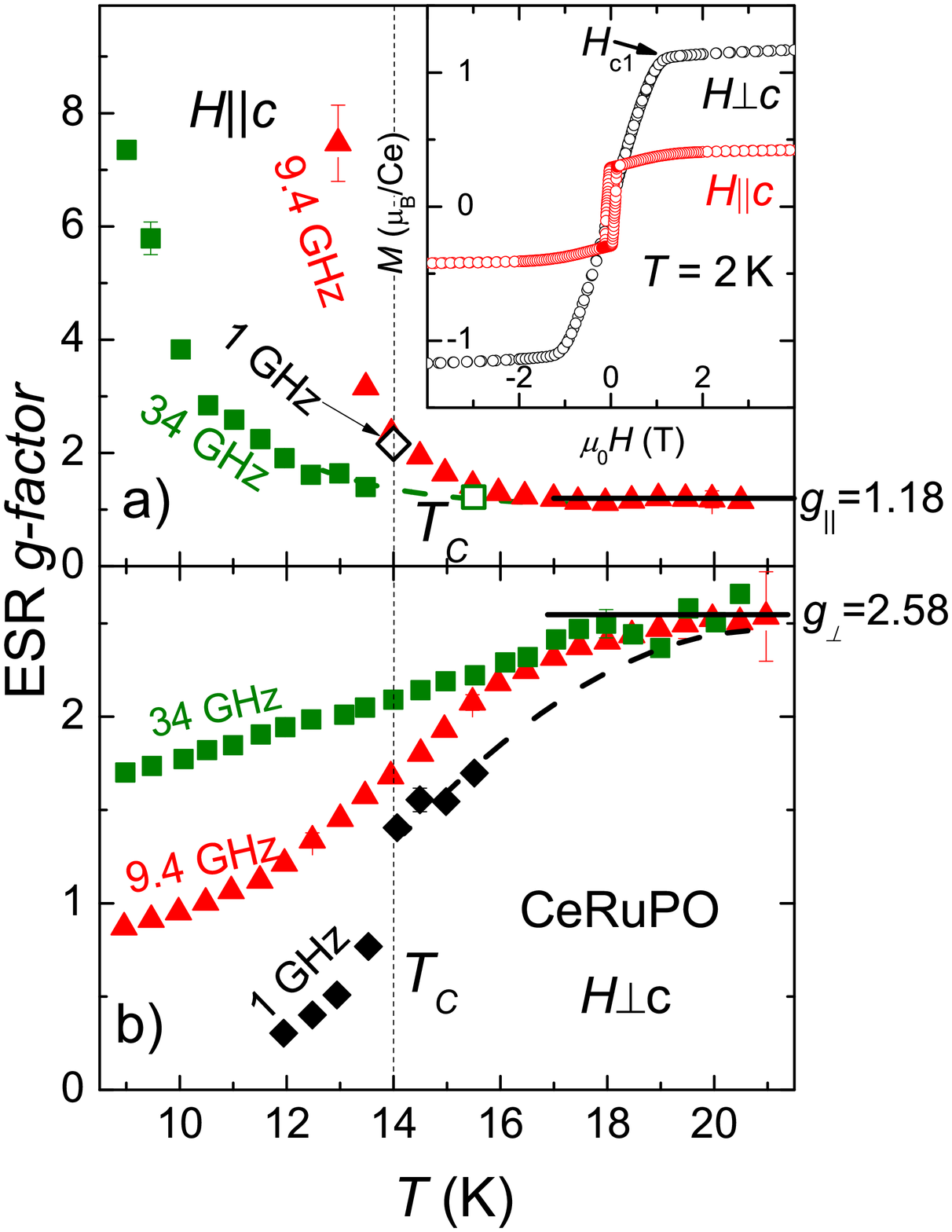}
 \caption{Temperature dependence of the effective ESR $g$-factors for a) $H\| c$  and b) $H\bot c$  at 1 \fulldiamond, 9.4 \textcolor{red}{\fulltriangle} and 34~GHz \textcolor[rgb]{0,0.5,0}{\fullsquare}. Dashed lines indicate an extrapolation towards the 9.4 GHz data. Data points for 34~GHz at 15.5~K $H\|c$ \textcolor[rgb]{0,0.5,0}{\opensquare} and for 1~GHz at 14~K $H\|c$ \opendiamond\ were extrapolated by fitting equation (\ref{gunixial}) to the angle dependence of the $g$-factor, see \autoref{gtheta}. Inset in a) shows the isothermal magnetization as function of the applied magnetic field at 2~K for $H\bot c$ \opencircle\ and $H\|c$ \textcolor{red}{\opencircle}, taken from Reference \cite{Krel08a}.}
\label{gtemp}
\end{figure}

Whether or not the observed resonance can be related to Ce$^{3+}$ magnetic moments could be checked by comparing the absolute values of $g_{\|}$ and $g_{\perp}$ in the paramagnetic region (see \autoref{gtemp}) with the $g$-values expected for Ce$^{3+}$ in a tetragonal CEF environment. There, the $J=\frac{5}{2}$ multiplet splits into three Kramers doublets with a single mixing coefficient $\eta$ \cite{Krel08a}:
\begin{eqnarray}
\Gamma_6\phantom{ ^{..\,}}&:& \left|\pm\frac{1}{2}\right\rangle \label{gamma6},\\
\Gamma_7 ^{(1)}&:& \eta \, \left| \pm\frac{5}{2}\right\rangle +\sqrt{1-\eta^2}\left|\mp\frac{3}{2}\right\rangle \label{gamma71},\\
\Gamma_7 ^{(2)}&:& \sqrt{1-\eta^2} \, \left|\pm\frac{5}{2}\right\rangle -\eta \left|\mp\frac{3}{2}\right\rangle .
\label{gamma72}
\end{eqnarray}
According to the saturation magnetization and the 4$f$-related specific heat, CeRuPO is likely to have a ground-state wave function with $\Gamma_6$ symmetry and two excited CEF levels at 70 and 320~K with a $\Gamma_7$ symmetry \cite{Krel08a}. 

If a $\Gamma_7$ ground-state is assumed and the admixture of excited CEF-doublets is neglected a calculation of the expected $g$-factors $g_{\bot,\|}^{th}$ \cite{Abra70} yields values that are inconsistent with the experimental data, regardless of which mixing coefficient $\eta$ is used. In contrast, with $g_{\bot}^{th}=$~2.57 and $g_{\|}^{th}=$~0.86 the $\Gamma_6$ doublet (\ref{gamma6}) provides results which agree well with the experimental values as shown in \autoref{gtemp}: the measured $g$-factors for $H\bot c$ and $H\|c$ become field and temperature independent above $T=18$~K, merging to $g_{\bot}$= 2.58 and $g_{\|}$= 1.18. Therefore, the observed ESR line can be associated with the resonance of a $\Gamma_6$ ground-state doublet of Ce$^{3+}$ in CeRuPO. The remaining differences between theoretical and experimental $g$-factors could originate from the commonly observed $g$-shift of paramagnetic ions in a metallic environment \cite{Barn81}.

%

The frequency dependence of $g_{\|}$ and $g_{\bot}$ displayed in \autoref{gtemp} corresponds to field-dependent internal magnetic fields which are determined by an interplay between the RKKY- and the CEF-interactions of the Ce$^{3+}$ magnetic moments. As mentioned above, the RKKY interaction leads to an FM ordering and alignment of the 4$f$-spins along the c-axis \cite{Krel08a}. Therefore, in the case $H\|c$, the internal and external fields sum up to the resonance field, i.e. one needs a much smaller external field to reach the resonance condition which corresponds to an increase of $g_{\|}$.


In the case of $H\bot c$ the $g$-factor continuously decreases with lowering the temperature across $T_C$. This decrease shows a frequency dependent slope that is largest for the lowest frequency. This behaviour originates from an effective FM coupling of the Ce$^{3+}$ magnetic moments along the $c$-axis which leads to a reduction of the magnetic field along the basal plane. Therefore a larger external magnetic field is necessary to reach the resonance condition and a decrease of $g_{\bot}$ is observed. The fact that this effect is largest at small frequencies (fields) can be related to the magnetic field dependence of the isothermal magnetization $M(H)$ of CeRuPO (see inset \autoref{gtemp}a)): When the magnetic field is applied along the crystallographic $c$-axis ($H\|c$), one observes a well defined hysteresis and a saturation with a magnetic moment of $\mu_{sat}^c=0.43\mu_B$. With the magnetic field applied perpendicular to the $c$-axis ($H\bot c$) no hysteresis is observed. Instead one finds a linear increase up to the critical field $\mu_0H_{c1}=1$~T and a saturated magnetic moment of $\mu_{sat}^{ab}=1.2\mu_B$ \cite{Krel08a}. For the ESR measured at 34~GHz, and $H\bot c$ the external field is swept between 1~T and 1.6~T which exceeds the saturation field $H_{\rm c1}$. Then, almost all FM moments are aligned along the basal plane, and the moment of the CEF ground state determines the $g_{\bot}$ factor. At frequencies where $H_{\rm c1}$ is larger than $H_{\rm res}$ (i.e. L- and X-band) the FM moments are not fully rotated towards the basal plane which, thus, results in a reduced $g_{\bot}$ factor.

%

\begin{figure}[t]
 \centering
\includegraphics[width=0.75\textwidth]{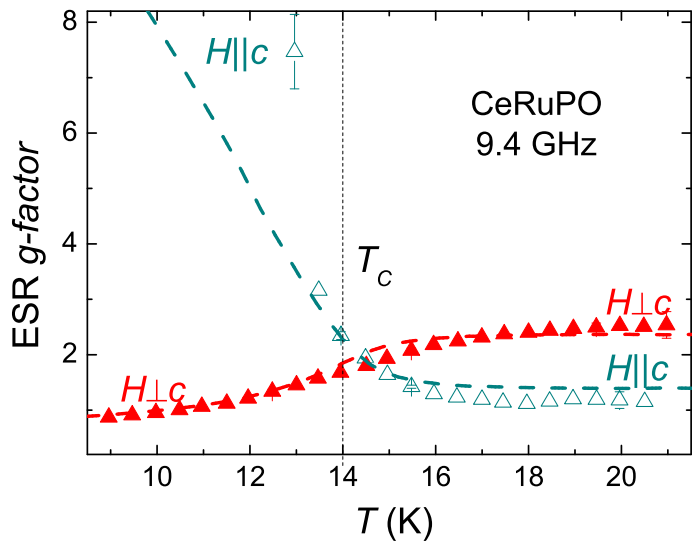}
 \caption{Temperature dependence of the effective ESR $g$-factor for $H\bot c$ \textcolor{red}{\fulltriangle} and $H\|c$ \textcolor[rgb]{0,0.5,0.5}{\opentriangle} at 9.4~GHz together with calculated $g$-factors using equations (\ref{hubergs}), (\ref{hubergp}) (dashed lines) and the susceptibility data measured at 0.1~T \cite{Krel08a}.}
\label{g-Temp+Huberg}
\end{figure}

A phenomenological approach to describe the temperature dependence of the $g$-factor in an anisotropic magnet has been realized by relating the $g$-factor to the bulk, static susceptibility \cite{Hube75}. This procedure has successfully been applied, for instance, for the uniaxial ferromagnet CrBr$_{3}$ and indicates that the ESR in concentrated systems probes a collective mode of the coupled spin system, in contrast 
to dilute systems where the resonance field is determined entirely by single-ion properties. For YbRh$_{2}$Si$_{2}$ and YbIr$_{2}$Si$_{2}$ the data could be well described by the following relations being valid if the magnetization is proportional to the applied field (``low-field limit'')  \cite{Hube09,Grun10}:
\begin{eqnarray}
g_{\bot}(T) &=& g^0_{\|} \sqrt{\frac{\chi_{\bot}(T)}{\chi_{\|}(T)}} \label{hubergs} \mbox{\hspace{0.5cm}and} \\
g_{\|}(T) &=& \frac{(g^{0}_{\bot})^2}{g^0_{\|}} \frac{\chi_{\|}(T)}{\chi_{\bot}(T)} \label{hubergp}.
\end{eqnarray}
Here, $g^0_{\|,\bot}$ belong to the microscopic $g$-tensor, and $\chi_{\|,\bot}$ denotes the susceptibility for $H\|c$ and $H\bot c$. The application of these relations to the ESR of CeRuPO yields reasonable results for the 9.4~GHz measurements as can be seen by the dashed lines in \autoref{g-Temp+Huberg}. For the descriptions of these data we used the susceptibility measured at a field of 0.1~T \cite{Krel08a} and $g^0_{\bot}=2.58$, $g^0_{\|}=1.18$ which are the values measured for $T>18$~K (see \autoref{gtemp}). We suspect that the deviations of the dashed lines from the data  mainly originate from the fact that a fixed field was used for the susceptibility measurements, while $\mu_0H_{res}$ changes from 0.3 to 1~T for $H\bot c$ and from 0.6 to 0.2~T for $H\|c$.

\subsection{Temperature and Frequency behaviour of the ESR linewidth}

\begin{figure}[ht]
 \centering
\includegraphics[width=0.6\textwidth]{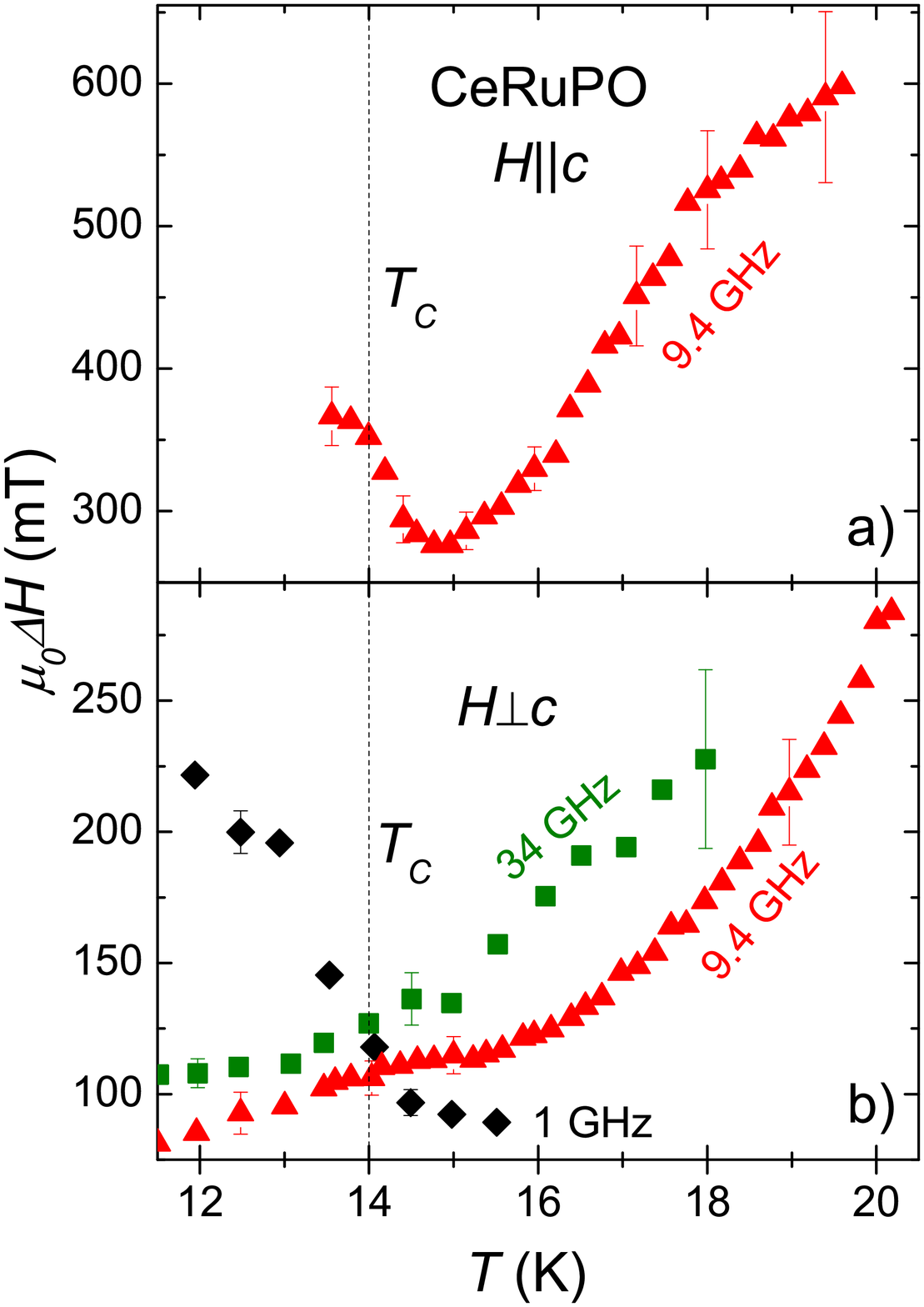}
 \caption{Temperature dependence of the ESR linewidth for a) $H\|c$  and b) $H\bot c$  at 1 \fulldiamond, 9.4 \textcolor{red}{\fulltriangle} and 34~GHz \textcolor[rgb]{0,0.5,0}{\fullsquare}.}
\label{dHTemp}
\end{figure}

The temperature and frequency behaviour of the linewidth of CeRuPO is plotted in \autoref{dHTemp}. $\Delta H_{\|}(T)$ could only be measured with the 9.4~GHz setup because of the lower sensitivity and larger linewidth at the other frequencies. It is worth noting the consistency with the linewidth data of the polycrystalline samples (see Fig. 2 in Ref. \cite{Krel08}): At temperatures above $T_C$ the linewidth of the polycrystals is dominated by the $H\bot c$ signal of the grains. Below $T_C$ the strong shift of the resonance field due the FM transition separates the signals, and the linewidth is dominated by the $H\|c$ component. 

In the paramagnetic regime the general linewidth behaviour resembles the common behaviour of \textit{diluted} $4f$-spins in a metallic environment \cite{Barn81}. There, a Korringa mechanism leads to a linear increase of the linewidth, which is then often followed by an exponential growth indicating the contribution of excited CEF levels. However, for our case of a high concentration of Ce$^{3+}$ ions in the presence of a Kondo effect the linewidth is determined by a more complicated mechanism. For example many important features of the ESR results of YbRh$_{2}$Si$_{2}$ could successfully be explained by the relaxation of a coupled spin mode of conduction electrons and Yb-$4f$ spins \cite{Koch09}.

In systems with concentrated magnetic moments the dipolar interaction can lead to a broadening of the linewidth. However, this effect can be neglected for CeRuPO  when considering a rough estimate for the high-temperature limit of the linewidth in presence of the exchange narrowing processes: $\Delta H^{\infty}\propto\frac{\left\langle\nu^2_{DD}\right\rangle}{\nu_{ex}}$. Here, $\left\langle\nu^2_{DD}\right\rangle$ is the second moment of the resonance-frequency distribution due to dipolar broadening \cite{Van48}, and $\nu_{ex}$ denotes the exchange frequency \cite{Ande53}. Although $\left\langle\nu^2_{DD}\right\rangle$ with inclusion of nearest and next-nearest neighbours results in linewidth values of the order of the experimental data, the exchange-narrowing process with $\nu_{ex}\approx \sqrt{zS(S+1)}I/h\ge 100$~GHz yields a reduction by at least two orders of magnitude. The exchange coupling $I$ was calculated using the Weiss-molecular equation $3k_B\Theta_{W}=IzS(S+1)$ ($z$: number of nearest and next-nearest neighbours, $\Theta_{W}$: Weiss temperatures taken from \cite{Krel08a} for both field orientations).

In the temperature region slightly above and below the magnetic ordering the linewidth data display remarkable dependencies on frequency and field orientation. For $H\bot c$ at 9.4~GHz and 34~GHz no anomaly is found around the onset of magnetic ordering. This contrasts to the data at 1~GHz ($H\bot c$) and at 9.4~GHz ($H\|c$) where the linewidth strongly increases upon decreasing the temperature across $T_{\rm C}$. These divergencies resemble the characteristics observed in various ferromagnets like Gd \cite{Burg77}, Ni \cite{Spoer75}, CrBr$_3$ \cite{Seeh74,Sche78} and CdCr$_2$Se$_4$ \cite{Kotz78}. They are referred to as "critical speeding-up" \cite{Kotz78} of the spin-relaxation time because  of a reduction of the exchange narrowing of magnetic dipole interactions when $T_{\rm C}$ is approached from above. This effect is present under the condition $H_{res} \ll H_{ex} \cdot (a/ \xi )^{5/2}$ which contains the exchange field $H_{ex}$, the lattice constant $a$ and the correlation length $\xi$ of the spin system \cite{Seeh74}.

In the case of CeRuPO this condition qualitatively may describe the presence or absence of divergent linewidth behavior near $T_{\rm C}$. The field dependence in \autoref{dHTemp}b shows that $H_{ex} \cdot (a/ \xi )^{5/2}$ is larger than the 1~GHz resonance field ($\approx 60$~mT) but similar or smaller than the resonance fields for the data at 9.4~GHz ($\approx 300$~mT) and 34~GHz ($\approx 1100$~mT). The orientational dependence in the 9.4~GHz data indicates that, when going from $H\| c$ to $H\bot c$, the critical fluctuations of the spontaneous magnetization become strongly suppressed, i.e. $H_{ex} \cdot (a/ \xi )^{5/2}$ gets reduced compared to $H_{res}$.

\section{Conclusion}
We presented a detailed study of ESR on CeRuPO single crystals at three different frequency bands. We have shown that the ESR signal displays the local properties of Ce$^{3+}$ ions with a $\Gamma_6$ CEF doublet ground-state in a metallic environment. Near $T_C$ the ESR is influenced by the critical fluctuations of the spontaneous magnetization, and at higher temperatures the magneto-crystalline anisotropy of CeRuPO dominates the resonance. 

The ESR of CeRuPO shares several peculiarities with the ESR of Yb-based Kondo-lattice systems YbRh$_2$Si$_2$ and YbIr$_2$Si$_2$, respectively: For instance, below the Kondo temperature all rare earth magnetic moments contribute to the resonance signal and, despite the presence of a \textit{dense} $4f$ lattice, anisotropy and temperature dependence of the resonance show a typical behaviour of \textit{diluted} 4$f$ moments in a metal. 

\section*{Acknowledgements}
We acknowledge the Volkswagen foundation (I/84689) for financial support.

\section*{References}
\bibliography{ESR_CeRuPO_EK}{}

\providecommand{\newblock}{}
\begin{thebibliography}{10}
\expandafter\ifx\csname url\endcsname\relax
  \def\url#1{{\tt #1}}\fi
\expandafter\ifx\csname urlprefix\endcsname\relax\def\urlprefix{URL }\fi
\providecommand{\eprint}[2][]{\url{#2}}

\bibitem{Sich03}
Sichelschmidt J, Ivanshin V~A, Ferstl J, Geibel C and Steglich F 2003 {\em
  Phys. Rev. Lett.\/} {\bf 91} 156401

\bibitem{Woelf09}
Wölfle P and Abrahams E 2009 {\em Phys. Rev. B\/} {\bf 80} 235112

\bibitem{Sich07}
Sichelschmidt J, Wykhoff J, Krug~von Nidda H~A, Fazlishanov I~I, Hossain Z,
  Krellner C, Geibel C and Steglich F 2007 {\em J. Phys.: Condes. Matter\/}
  {\bf 19} 016211

\bibitem{Krel08}
Krellner C, Förster T, Jeevan H, Geibel C and Sichelschmidt J 2008 {\em Phys.
  Rev. Lett.\/} {\bf 100} 066401

\bibitem{Krel07a}
Krellner C, Kini N~S, Brüning E~M, Koch K, Rosner H, Nicklas M, Baenitz M and
  Geibel C 2007 {\em Phys. Rev. B\/} {\bf 76} 104418

\bibitem{Schl09}
Schlottmann P 2009 {\em Phys. Rev. B\/} {\bf 79} 045104

\bibitem{Koch09}
Kochelaev B~I, Belov S~I, Skvortsova A~M, Kutuzov A~S, Sichelschmidt J, Wykhoff
  J, Geibel C and Steglich F 2009 {\em Eur. Phys. J. B\/} {\bf 72} 485--489

\bibitem{Zvya09a}
Zvyagin A~A, Kataev V and B\"uchner B 2009 {\em Phys. Rev. B\/} {\bf 80} 024412

\bibitem{Huan76}
Huang C~Y, K S and Cooper B~R 1976 {\em Proc. Int. Conf. Cryst. Field Effects
  Met. Alloys\/} {\bf -} 51--60

\bibitem{Demi09a}
Demishev S~V, Semeno A~V, Bogach A~V, Samarin N~A, Ishchenko T~V, Filipov V~B,
  Shitsevalova N~Y and Sluchanko N~E 2009 {\em Phys. Rev. B\/} {\bf 80} 245106

\bibitem{Krel08a}
Krellner C and Geibel C 2008 {\em J. Cryst. Growth\/} {\bf 310} 1875--1880

\bibitem{Wykh07a}
Wykhoff J, Sichelschmidt J, Lapertot G, Knebel G, Flouquet J, Fazlishanov I~I,
  von Nidda H~A~K, Krellner C, Geibel C and Steglich F 2007 {\em Science And
  Technology Of Advanced Materials\/} {\bf 8} 389--392

\bibitem{Anis99}
Anisimov A~N, Farle M, Poulopoulos P, Platow W, Baberschke K, Isberg P,
  W\"appling R, Niklasson A~M~N and Eriksson O 1999 {\em Phys. Rev. Lett.\/}
  {\bf 82} 2390--2393

\bibitem{Farl00}
Farle M, Anisimov A~N, K B, J L and H M 2000 {\em Europhys. Lett.\/} {\bf 49}
  658--664

\bibitem{Kotz75}
Kötzler J and Scheithe W 1975 {\em Phys. Status Solidi B\/} {\bf 69} 389--394

\bibitem{Abra70}
Abragam A and Bleaney B 1970 {\em Electron Paramagnetic Resonance of Transition
  Ions\/} (Oxford University Press)

\bibitem{Barn81}
Barnes S~E 1981 {\em Adv. Phys.\/} {\bf 30} 801--938

\bibitem{Hube75}
Huber D~L and Seehra M~S 1975 {\em J. Phys. Chem. Solids\/} {\bf 36} 723--725

\bibitem{Hube09}
Huber D~L 2009 {\em Journal of Physics: Condensed Matter\/} {\bf 21} 322203

\bibitem{Grun10}
Gruner T, Wykhoff J, Sichelschmidt J, Krellner C, Geibel C and Steglich F 2010
  {\em Journal of Physics: Condensed Matter\/} {\bf 22} 135602

\bibitem{Van48}
Van~Vleck J~H 1948 {\em Phys. Rev.\/} {\bf 74} 1168--1183

\bibitem{Ande53}
Anderson P~W and Weiss P~R 1953 {\em Rev. Modern Phys.\/} {\bf 25} 269--276

\bibitem{Burg77}
Burgardt P and Seehra M~S 1977 {\em Phys. Rev. B\/} {\bf 16} 1802--1807

\bibitem{Spoer75}
Spörel F and Biller E 1975 {\em Solid State Commun.\/} {\bf 17} 833 -- 835

\bibitem{Seeh74}
Seehra M~S and Gupta R~P 1974 {\em Phys. Rev. B\/} {\bf 9} 197--202

\bibitem{Sche78}
Scheithe W, Kötzler J and Radhakrishna P 1978 {\em Phys. Lett. A\/} {\bf 66}
  419--421

\bibitem{Kotz78}
Kötzler J and Philipsborn H~V 1978 {\em Phys. Rev. Lett.\/} {\bf 40} 790--793

\end{thebibliography}
\bibliographystyle{iopart-num-without-url}

\end{document}